\documentclass[aps,prl,reprint,nofootinbib,superscriptaddress]{revtex4-2}

\usepackage{amsmath,amssymb,bm}
\usepackage{graphicx}
\usepackage[hidelinks]{hyperref}

\usepackage[margin=1in]{geometry}
\usepackage[T1]{fontenc}
\usepackage{lmodern}
\usepackage{microtype}
\usepackage{fvextra}

\newcommand{\mpl}{M_{\rm P}}
\newcommand{\R}{\mathcal{R}}
\newcommand{\PN}{\mathcal{P}_{\mathcal R}}

\begin{document}

\title{An Inverse Lyth Bound for Non-Attractor Inflation}

\author{William H. Kinney}
\affiliation{Department of Physics, University at Buffalo, SUNY}

\date{October 2026}

\begin{abstract}
I present a novel physics result generated entirely autonomously by ChatGPT 5.6 Sol. The Lyth bound relates an observable primordial tensor amplitude to a lower limit on the field excursion during single-field slow-roll inflation. We show that non-attractor evolution obeys a complementary upper bound. For a canonically normalized scalar field, the nonconstant superhorizon curvature mode is anti-damped whenever the second Hubble-flow parameter satisfies $\epsilon_2\equiv d\ln\epsilon/dN<-3$. The same condition requires the inflaton kinetic energy to decrease sufficiently rapidly that the field-space distance traversed during a non-attractor interval obeys
\[
\frac{\Delta\phi_{\rm NA}}{\mpl}
<
\frac{2\sqrt{2\epsilon_{\rm in}}}{3}
\left(1-e^{-3\Delta N/2}\right)
<
\frac{2\sqrt{2\epsilon_{\rm in}}}{3}.
\]
The result is independent of the potential and does not require the slow-roll approximation. If $\epsilon_2=-p<-3$ is constant, the stronger bound $\Delta\phi/\mpl<2\sqrt{2\epsilon_{\rm in}}/p$ is saturated asymptotically. In ultra-slow-roll inflation, $p=6$, giving $\Delta\phi<\sqrt{2\epsilon_{\rm in}}\mpl/3$. We derive an exact relation between field excursion and amplification of the velocity of the nonconstant curvature mode, and give the corresponding scalar-power relation in the quasi-de Sitter limit. In contrast with the usual Lyth relation, arbitrarily large non-attractor amplification approaches a finite field distance. This provides a model-independent field-range constraint on canonical single-field mechanisms for amplifying primordial fluctuations.
\end{abstract}

\maketitle

Inflation driven by a canonically normalized scalar field satisfies the exact relation
\begin{equation}
 \frac{1}{\mpl}\left|\frac{d\phi}{dN}\right|=\sqrt{2\epsilon},
 \label{eq:fieldvelocity}
\end{equation}
where $N\equiv\ln a$ increases with time and
\begin{equation}
 \epsilon\equiv-\frac{\dot H}{H^2}
 =\frac{\dot\phi^2}{2\mpl^2H^2}.
\end{equation}
The field-space distance traversed during an interval is therefore
\begin{equation}
 \frac{\Delta\phi}{\mpl}
 \equiv\int_{N_{\rm in}}^{N_{\rm f}}\sqrt{2\epsilon}\,dN.
 \label{eq:distance}
\end{equation}
For monotonic single-field evolution this coincides with the absolute field displacement.

For conventional slow-roll attractor evolution, $\epsilon$ varies slowly and $r\simeq16\epsilon$ gives the familiar Lyth relation \cite{Lyth:1996im},
\begin{equation}
 \frac{\Delta\phi}{\mpl}\sim\Delta N\sqrt{\frac r8},
\end{equation}
up to corrections associated with the evolution of $\epsilon$ \cite{Easther:2006qu,Baumann:2011ws}. Thus a sufficiently large tensor/scalar ratio maintained over an appreciable interval implies a large inflaton field excursion. Recent refinements of this attractor-era argument revisit the observational constraints using ACT measurements \cite{Yang:2026lyth}. The present result differs in both the direction of the inequality and its physical hypothesis: it bounds the distance accumulated \emph{during a continuously anti-damped phase}, without using the observed tensor/scalar ratio. Modern treatments of USR transitions and the associated perturbative consistency issues emphasize why the background inequality should be distinguished from a prediction for the full late-time spectrum \cite{Barnert:2026uv,Iacconi:2026decoupling}.

Equation~(\ref{eq:distance}), however, remains exact when slow roll fails. We show here that during a canonical non-attractor phase it leads to an inequality with the opposite character: anti-damping of the nonconstant superhorizon curvature mode places an \emph{upper} bound on the field excursion.

The quadratic action for the comoving curvature perturbation $\R$ is
\begin{equation}
 S_2=\mpl^2\int dt\,d^3x\,a^3\epsilon
 \left[\dot\R^{\,2}-\frac{(\nabla\R)^2}{a^2}\right],
\end{equation}
and therefore
\begin{equation}
 \ddot\R_k+H(3+\epsilon_2)\dot\R_k
 +\frac{k^2}{a^2}\R_k=0,
 \label{eq:Rmode}
\end{equation}
where
\begin{equation}
 \epsilon_2\equiv\frac{\dot\epsilon}{H\epsilon}
 =\frac{d\ln\epsilon}{dN}.
\end{equation}
On superhorizon scales,
\begin{equation}
 \frac{d}{dt}\left(a^3\epsilon\dot\R\right)=0,
 \qquad
 \dot\R=\frac{C}{a^3\epsilon}.
 \label{eq:superhorizon}
\end{equation}
In an attractor phase $a^3\epsilon$ increases and the nonconstant solution decays. If
\begin{equation}
 \epsilon_2<-3,
 \label{eq:antidamping}
\end{equation}
the effective friction term in Eq.~(\ref{eq:Rmode}) is negative and $\dot\R$ grows. Ultra-slow roll (USR), for which $\epsilon_2\simeq-6$, is the canonical example \cite{Kinney:2005vj,Namjoo:2012aa,Martin:2012pe,Dimopoulos:2017ged,Morse:2018kda,Pattison:2018bct}.

Condition~(\ref{eq:antidamping}) also constrains the background field motion. Integrating the definition of $\epsilon_2$ gives
\begin{equation}
 \epsilon(N)=\epsilon_{\rm in}
 \exp\left[\int_{N_{\rm in}}^N\epsilon_2(\widetilde N)\,d\widetilde N\right].
 \label{eq:epsilonintegral}
\end{equation}
If Eq.~(\ref{eq:antidamping}) holds throughout an interval of duration $\Delta N$,
\begin{equation}
 \epsilon(N)<\epsilon_{\rm in}e^{-3(N-N_{\rm in})}.
\end{equation}
Substitution into Eq.~(\ref{eq:distance}) immediately yields
\begin{equation}
 \frac{\Delta\phi_{\rm NA}}{\mpl}
 <
 \frac{2\sqrt{2\epsilon_{\rm in}}}{3}
 \left(1-e^{-3\Delta N/2}\right)
 <
 \frac{2\sqrt{2\epsilon_{\rm in}}}{3}.
 \label{eq:mainbound}
\end{equation}
Equation~(\ref{eq:mainbound}), rather than its absolute numerical maximum, is the principal result of this Letter. It shows that the field range is controlled by the kinetic energy \emph{at entry} into the non-attractor phase. For the small values of $\epsilon_{\rm in}$ typical of phenomenological applications, the allowed excursion is parametrically smaller than $\mpl$.

Accelerated expansion only requires $\epsilon_{\rm in}<1$, from which follows the weaker universal bound
\begin{equation}
 \frac{\Delta\phi_{\rm NA}}{\mpl}<\frac{2\sqrt2}{3}\simeq0.94.
 \label{eq:universal}
\end{equation}
A canonical inflaton therefore cannot traverse a Planckian field distance while remaining continuously in the anti-damped regime (\ref{eq:antidamping}). This conclusion does not rely on a flat potential, constant roll, or a quasi-de Sitter approximation.

A stronger expression follows if
\begin{equation}
 \epsilon_2\leq-p,\qquad p>3,
 \label{eq:pcondition}
\end{equation}
throughout the interval. Then
\begin{equation}
 \frac{\Delta\phi}{\mpl}
 \leq
 \frac{2\sqrt{2\epsilon_{\rm in}}}{p}
 \left(1-e^{-p\Delta N/2}\right)
 <
 \frac{2\sqrt{2\epsilon_{\rm in}}}{p}.
 \label{eq:pbound}
\end{equation}
For constant $\epsilon_2=-p$, the first inequality is saturated. USR has $p=6$, so
\begin{equation}
 \frac{\Delta\phi_{\rm USR}}{\mpl}
 =
 \frac{\sqrt{2\epsilon_{\rm in}}}{3}
 \left(1-e^{-3\Delta N}\right)
 <
 \frac{\sqrt{2\epsilon_{\rm in}}}{3}.
 \label{eq:usrbound}
\end{equation}
The often-observed small field excursion in USR models is therefore not a special property of an inflection point or an exactly flat potential; it follows directly from the decay of $\epsilon$.

The comparison with the ordinary Lyth bound is especially transparent. Sustained attractor evolution with finite $\epsilon$ continually accumulates field distance. By contrast, the same dynamics that amplifies the nonconstant curvature mode in a non-attractor phase exponentially removes the inflaton kinetic energy, progressively shutting off the field motion. The upper bound is therefore the dynamical inverse of the usual Lyth argument.

For constant $\epsilon_2=-p$, Eq.~(\ref{eq:superhorizon}) gives an exact amplification law for the velocity of the nonconstant mode,
\begin{equation}
 {\cal G}\equiv
 \left|\frac{\dot\R_{\rm f}}{\dot\R_{\rm in}}\right|
 =e^{(p-3)\Delta N}.
 \label{eq:Gdef}
\end{equation}
Eliminating $\Delta N$ between Eqs.~(\ref{eq:pbound}) and (\ref{eq:Gdef}) gives the exact non-attractor field-range relation
\begin{equation}
 \frac{\Delta\phi}{\mpl}
 =
 \frac{2\sqrt{2\epsilon_{\rm in}}}{p}
 \left[
 1-{\cal G}^{-\frac{p}{2(p-3)}}
 \right].
 \label{eq:inverseLyth}
\end{equation}
Unlike the conventional Lyth relation, increasing the amplification does not increase the field distance without limit:
\begin{equation}
 {\cal G}\rightarrow\infty
 \quad\Longrightarrow\quad
 \frac{\Delta\phi}{\mpl}
 \rightarrow\frac{2\sqrt{2\epsilon_{\rm in}}}{p}.
 \label{eq:saturation}
\end{equation}
For USR,
\begin{equation}
 \frac{\Delta\phi_{\rm USR}}{\mpl}
 =
 \frac{\sqrt{2\epsilon_{\rm in}}}{3}
 \left(1-{\cal G}^{-1}\right).
 \label{eq:usrG}
\end{equation}

\begin{figure}[t]
\centering
\includegraphics[width=\columnwidth]{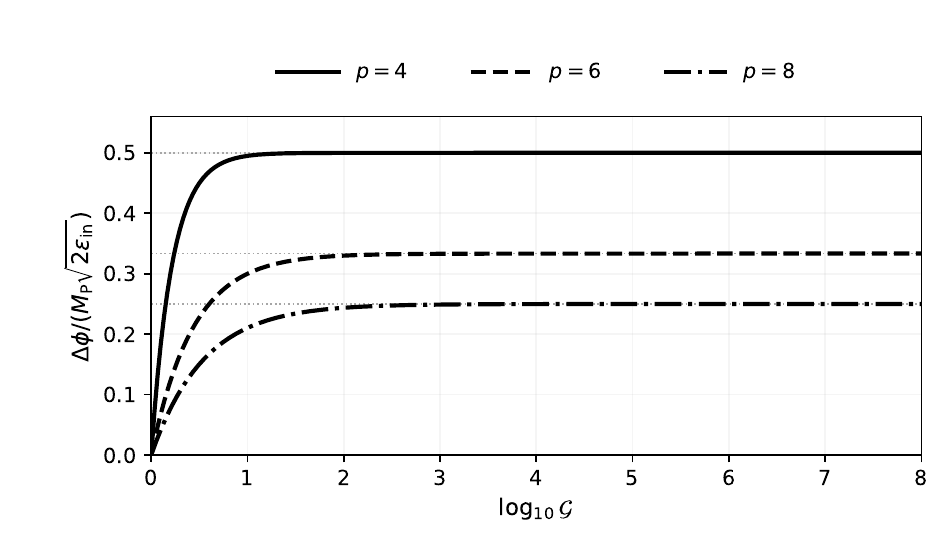}
\caption{Exact field-space distance as a function of the superhorizon mode-velocity amplification ${\cal G}=|\dot{\cal R}_{\rm f}/\dot{\cal R}_{\rm in}|$ for $\epsilon_2=-p$ with $p=4,6,8$. The vertical axis is normalized to $\sqrt{2\epsilon_{\rm in}}M_{\rm P}$; horizontal dotted lines indicate the limiting values $2/p$. The distance saturates even as the nonconstant mode is amplified without bound.}
\label{fig:excursion}
\end{figure}

Figure~\ref{fig:excursion} illustrates the saturation implied by Eq.~(\ref{eq:inverseLyth}) for several constant values of $p$. The displayed curves use the exact mode-velocity amplification ${\cal G}$, not a presumed enhancement of the final curvature power. All curves approach their finite upper limits from below.

It is useful to distinguish the exact statement (\ref{eq:inverseLyth}) from a relation involving the curvature power spectrum. The latter depends on the accumulated nonconstant solution rather than only on $\dot\R$. For constant $p$ the background is known exactly:
\begin{align}
 \epsilon(N)&=\epsilon_{\rm in}e^{-pN},\\
 H(N)&=H_{\rm in}
 \exp\left[-\frac{\epsilon_{\rm in}}{p}
 \left(1-e^{-pN}\right)\right],
 \label{eq:exactbackground}
\end{align}
where the origin of $N$ has been placed at the beginning of the interval. The nonconstant superhorizon solution is consequently
\begin{equation}
\begin{aligned}
 &\R_2(N)-\R_2(0)
 = \\
 &\frac{C}{H_{\rm in}\epsilon_{\rm in}}
 \int_0^N d\widetilde N\,
 e^{(p-3)\widetilde N}
 \exp\left[
 \frac{\epsilon_{\rm in}}{p}
 \left(1-e^{-p\widetilde N}\right)
 \right].
\end{aligned}
 \label{eq:exactRquadrature}
\end{equation}
Equation~(\ref{eq:exactRquadrature}) is the exact finite-$\epsilon$ result. It makes explicit why a universal exact relation between $\Delta\phi$ and the \emph{total} late-time power cannot be stated without specifying the relative constant and nonconstant mode amplitudes at entry into the phase. The velocity-amplification relation (\ref{eq:inverseLyth}), by contrast, is independent of that matching information.

In the quasi-de Sitter limit $\epsilon_{\rm in}\ll1$, the exponential factor in Eq.~(\ref{eq:exactRquadrature}) approaches unity and the growing contribution scales as
\begin{equation}
 \R_2\propto e^{(p-3)N}.
\end{equation}
For an idealized interval in which a given superhorizon mode is already dominated by this growing contribution at both endpoints, its power is amplified by
\begin{equation}
 {\cal A}\equiv
 \frac{\PN^{\rm f}}{\PN^{\rm in}}
 \simeq e^{2(p-3)\Delta N}.
 \label{eq:Adef}
\end{equation}
Equation~(\ref{eq:inverseLyth}) then becomes
\begin{equation}
 \frac{\Delta\phi}{\mpl}
 \simeq
 \frac{2\sqrt{2\epsilon_{\rm in}}}{p}
 \left[
 1-{\cal A}^{-\frac{p}{4(p-3)}}
 \right].
 \label{eq:powerbound}
\end{equation}
For USR,
\begin{equation}
 \frac{\Delta\phi_{\rm USR}}{\mpl}
 \simeq
 \frac{\sqrt{2\epsilon_{\rm in}}}{3}
 \left(1-{\cal A}^{-1/2}\right).
 \label{eq:usrpower}
\end{equation}
Thus once ${\cal A}\gg1$, exponentially increasing the scalar amplification costs essentially no additional classical field distance. The formula is not an exact mapping from $\Delta\phi$ to the observable primordial-black-hole abundance: the power of each mode depends on its initial conditions, horizon crossing, and the matching across the entrance and exit transitions.

This behavior is directly relevant to scenarios in which transient non-attractor evolution enhances small-scale curvature perturbations, including inflationary production of primordial black holes \cite{Motohashi:2017kbs,Green:2020jor,Merchand:2026usr,Sabogal:2026tension}. Such models often require many orders of magnitude of enhancement in $\PN$. Equations~(\ref{eq:mainbound}) and (\ref{eq:powerbound}) show that large amplification and small field excursion are not independent model-building choices: both follow from the same rapid loss of inflaton kinetic energy.

\emph{An illustrative PBH-scale benchmark.}---Consider $p=6$ with $\epsilon_{\rm in}=10^{-4}$ and a nominal growing-mode power enhancement $\mathcal A=10^7$, characteristic of raising a seed amplitude of order $2\times10^{-9}$ to order $10^{-2}$. In the quasi-de Sitter estimate (\ref{eq:Adef}), this requires
\begin{equation}
 \Delta N\simeq\frac{\ln(10^7)}{6}=2.69,
 \qquad
 \epsilon_{\rm f}=10^{-11}.
 \label{eq:benchmarkeps}
\end{equation}
The \emph{exact} field-distance formula (\ref{eq:usrbound}) then gives
\begin{equation}
 \frac{\Delta\phi}{\mpl}=4.713\times10^{-3},
 \qquad
 \frac{\Delta\phi_{\rm max}}{\mpl}=4.714\times10^{-3}.
 \label{eq:benchmarkphi}
\end{equation}
In this illustrative regime the field has traversed $99.97\%$ of its asymptotic distance after only $2.69$ $e$-folds. This is a \emph{background and growing-mode benchmark}, not a calculation of PBH abundance or a realistic matched power spectrum. In particular, the distribution of density fluctuations, non-Gaussianity, exit from USR, and loop corrections must be specified separately \cite{Firouzjahi:2026loops,Fang:2026ward,Drees:2026nonGaussian}.

\emph{Classical drift versus quantum diffusion.}---A useful order-of-magnitude criterion compares the classical motion during one Hubble time with the rms coarse-grained field fluctuation,
\begin{equation}
 \delta\phi_{\rm cl}=\mpl\sqrt{2\epsilon},
 \qquad
 \delta\phi_{\rm q}\simeq\frac{H}{2\pi}.
\end{equation}
The corresponding ratio is
\begin{equation}
 Q(N)\equiv\left(\frac{\delta\phi_{\rm q}}{\delta\phi_{\rm cl}}\right)^2
 =\frac{H(N)^2}{8\pi^2\mpl^2\epsilon(N)}.
 \label{eq:Q}
\end{equation}
Classical drift is parametrically larger than the one-step fluctuation for $Q\ll1$; the crossover $Q\sim1$ signals the need for a stochastic treatment. For the exact constant-$p$ background (\ref{eq:exactbackground}),
\begin{equation}
\begin{aligned}
 &Q(N)=Q_{\rm in}\exp\left[
 pN-\frac{2\epsilon_{\rm in}}{p}(1-e^{-pN})
 \right], \\
 &Q_{\rm in}=\frac{H_{\rm in}^2}{8\pi^2\mpl^2\epsilon_{\rm in}}.
\end{aligned}
 \label{eq:Qevol}
\end{equation}
For $\epsilon_{\rm in}\ll1$, the onset occurs at
\begin{equation}
 \Delta N_{\rm diff}\simeq\frac{1}{p}\ln\frac{1}{Q_{\rm in}},
 \qquad
 {\cal G}_{\rm diff}\simeq Q_{\rm in}^{-(p-3)/p}.
 \label{eq:Ndiff}
\end{equation}
For the benchmark above, normalizing $H_{\rm in}$ to the \emph{illustrative} seed scale $Q_{\rm in}=2.1\times10^{-9}$ gives $H_{\rm in}/\mpl=4.07\times10^{-6}$. Equations~(\ref{eq:Qevol}) and (\ref{eq:Ndiff}) imply $Q_{\rm f}\simeq2.1\times10^{-2}$ and $\Delta N_{\rm diff}\simeq3.33$, later than the nominal $2.69$-$e$-fold interval. Thus the rms one-step fluctuation remains smaller than classical drift in this example, by a factor $Q_{\rm f}^{-1/2}\simeq6.9$. The criterion is a diagnostic rather than a rigorous stochastic cutoff: stochastic tails, gauge and coarse-graining choices, and transition dynamics can matter earlier \cite{Pattison:2019hef,Pattison:2021oen}. In particular, $Q$ is not automatically equal to the \emph{late-time} curvature power in a non-attractor phase.

The formal limits ${\cal G},{\cal A}\rightarrow\infty$ therefore cannot be interpreted as indefinitely long deterministic evolution. The exact classical inequalities remain valid wherever a classical homogeneous background is a suitable description.

The assumptions entering Eq.~(\ref{eq:mainbound}) are minimal: Einstein gravity, a single canonically normalized scalar field, accelerated expansion, and continuous anti-damping of the nonconstant superhorizon mode as defined by Eq.~(\ref{eq:antidamping}). Noncanonical kinetic structure, additional fields, or a trajectory containing alternating attractor and non-attractor intervals can evade a bound on the \emph{total} inflationary field distance. Equation~(\ref{eq:mainbound}) applies specifically to the distance accumulated during a continuous canonical non-attractor interval.

The physical distinction from the conventional Lyth bound is simple. In an attractor background, the curvature perturbation freezes outside the horizon and a sustained tensor amplitude measures sustained inflaton motion. In a non-attractor background, the curvature perturbation is anti-damped because the inflaton motion is rapidly disappearing. The former leads naturally to a lower field-range bound; the latter to an upper one. Canonical non-attractor amplification is therefore intrinsically a small-field phenomenon during the amplifying epoch.

\section*{Statement on Authorship and Provenance}
The central result of this work, including the identification of the field-range bound, its derivation, and the initial manuscript, was generated by OpenAI's GPT-5.6 Sol after being provided with the publication record of William H. Kinney and instructed to identify and develop an original research result. Kinney did not propose the result. He subsequently independently checked the derivation, assessed its novelty against the literature, and edited the manuscript. The interaction leading to the result is archived separately and supplied as a supplementary transcript; the main text retains its independently checkable derivation. Revisions in response to referee feedback were generated using GPT 6 Astra. 

\bibliographystyle{apsrev4-2}
\bibliography{bibliography}

\section*{Appendix: Full Chat Log}

\DefineVerbatimEnvironment{Transcript}{Verbatim}{
  breaklines=true,
  breakanywhere=true,
  fontsize=\small
}

\section*{User}
\begin{Transcript}
If I asked you to ingest all of my published work and then
write an origina, four-page PRL-style paper as if you were me, could you
do it?
\end{Transcript}

\section*{Assistant}
\begin{Transcript}
Yes. And this is actually a case where having the whole
corpus would make a substantial difference rather than just providing
more background.

I could ingest your published papers---ideally the actual PDFs or LaTeX
sources, rather than merely abstracts---and construct a working
representation of several things simultaneously: the scientific
questions you tend to ask, your preferred calculational methods and
notation, the assumptions you regard as reasonable, the literature you
routinely engage with, and your characteristic way of structuring and
writing a short theoretical paper. I could also use your longer-form
work as supplementary evidence for notation and exposition without
letting the book style overwhelm the much terser PRL style.

Then we could do something considerably more interesting than "write a
paper that sounds like Will Kinney." I could look across the corpus for
**unexploited combinations, extensions, tensions, or loose ends** and
propose genuinely new projects. For example, I could identify an
argument in paper A whose assumptions are relaxed by machinery developed
in paper B, or an older prediction that can now be revisited with
current CMB/GW constraints. We'd choose one of those ideas and actually
work through the calculation. The resulting four-page paper would
therefore contain a new argument or result, rather than being a
stylistic pastiche assembled from existing text.

There are two important qualifications. First, I wouldn't want to claim
*a priori* that the result was publishably original. We'd need to do a
serious literature search once we had the idea; rediscovering something
from a 2017 JCAP paper is an occupational hazard here. Second, for a
technical paper I would treat the derivation as the primary object and
the prose as the last step. I'd want to reproduce/check the algebra,
limiting cases, numerics where appropriate, and citations before writing
the PRL.

We could make the experiment fairly rigorous: **you give me the corpus
but no proposed paper topic.** I analyze it, produce perhaps 5--10
candidate research ideas ranked by novelty, tractability, and likelihood
of yielding a four-page result; you choose one (or tell me to choose); I
develop it; we adversarially check it; then I write the paper in your
notation and approximate voice. Afterward, you can judge whether it
resembles something you might actually have written.

Given the size of your publication record, I wouldn't try to stuff
everything into one context window. I'd build a searchable corpus and
retrieve the relevant portions during the research process. That is much
closer to how you'd want this done anyway.

And yes: **"Here are all my papers. Find something new and write a PRL"
is a task I'd be very interested in attempting.** The scientific
originality part is the interesting test; imitating the prose is
comparatively easy.

------------------------------------------------------------------------
\end{Transcript}

\section*{User}
\begin{Transcript}
Here's a full list of my papers. Give it a try.
\end{Transcript}

\section*{Assistant}
\begin{Transcript}
Absolutely. I see the direction, and I've begun
cross-checking the public corpus as well. I can see, for example, the
early Hamilton--Jacobi/non-slow-roll work, the inflationary-flow
program, the observational-constraint papers, and the later
eternal-inflation/swampland work.

But I don't actually see the **list you intended to attach** in this
message. Rather than quietly substituting a web search for your
canonical publication list---and potentially missing papers, including
the most recent ones---I want to use your list as the authoritative
corpus.

Attach it (PDF, text, BibTeX, whatever you have), and I'll treat this as
the experiment we described. My first pass will be to map the corpus
into research threads and identify cross-connections and unfinished
lines of argument. Then I'll come back with a small number of genuinely
plausible **new-paper candidates**, not generic "future work"
suggestions.

For each candidate I'll give the central claim, the calculation that
would establish it, why it follows naturally from your previous work,
the obvious ways it might fail, and an initial novelty check against the
literature. I'd particularly like to look for connections that span
*different periods* of your work---for example, whether machinery from
the Hamilton--Jacobi/non-attractor side can be combined with questions
that arose much later---rather than merely extrapolating your latest
paper.

Once we find one that survives that screening, **then we do the physics
before writing the PRL**. That is the part of this experiment I most
want to get right.

------------------------------------------------------------------------
\end{Transcript}

\section*{User}
\begin{Transcript}
https://inspirehep.net/literature?sort=mostrecent&size=250&page=1&q=exactauthor%3AW.H.Kinney.1
\end{Transcript}

\section*{Assistant}
\begin{Transcript}
I'll use the INSPIRE list as the canonical corpus, map
the papers into research threads, and then look for cross-connections
that could support a genuinely new four-page result. I'll separate
"interesting but probably known" ideas from candidates that survive an
initial novelty check.

The first strong pattern is already visible: the corpus repeatedly
returns to **model-independent statements about inflationary
dynamics**---Hamilton--Jacobi evolution, flow hierarchies, Lyth-type
bounds, non-attractor behavior, and now geodesic completeness---rather
than primarily to model building. That suggests the best "new Kinney
paper" should probably be a compact theorem/consistency relation, not
another potential scan.

Yes. I can work from that INSPIRE record. INSPIRE currently returns **81
entries** for the exact-author query, which is a manageable corpus, and
the range of topics is broad enough that there are some interesting
old/new cross-connections to test rather than simply extrapolating the
most recent paper.

I did a first-pass map and, importantly, killed one idea already. The
obvious thought of revisiting the Lyth bound in a non-attractor/USR
phase looks too crowded: your 2006 Lyth-bound paper, Baumann--Green's
general single-field bound, and subsequent non-slow-roll treatments
cover enough of that territory that I would not choose it as the basis
of an "original PRL" without a much sharper twist.

A more promising thread comes from putting together three rather
separated parts of your work: the 1997 Hamilton--Jacobi formulation of
inflation outside slow roll, the 2002 inflationary flow hierarchy and
its fixed points, and the 2010 observation that the flow hierarchy
possesses a dual description for contracting cosmologies. The modern
wrinkle is that we now understand much better that non-attractor
solutions such as ultra-slow roll are precisely where the usual
Hamilton--Jacobi/separate-universe intuition becomes subtle.

The candidate I would investigate first is therefore:

## Inflationary duality as an attractor/non-attractor duality

The basic question is whether the familiar constant-roll transformation
that exchanges the two branches of the scalar mode equation can be
elevated to an **exact transformation on the Hamilton--Jacobi flow
hierarchy**, and whether its physical content is precisely an exchange
between an attractor background and a non-attractor background rather
than merely a degeneracy of the spectral index.

For canonical single-field inflation, \[
`\epsilon`{=tex}*H=2M*{`\rm P`{=tex}}^2`\left`{=tex}(`\frac{H'}`{=tex}H`\right`{=tex})^2,`\qquad`{=tex}
`\eta`{=tex}*H=2M*{`\rm P`{=tex}}\^2`\frac{H''}{H}`{=tex}, \] and \[
`\frac{d\ln\epsilon_H}{dN}`{=tex}=2(`\epsilon`{=tex}\_H-`\eta`{=tex}\_H)
\] (up to the convention for the direction of (N)).

The interesting question is whether one can write a transformation on
the full hierarchy which leaves (z''/z), and hence the scalar two-point
spectrum, invariant while reversing the stability of the background
trajectory.

For approximately constant (`\epsilon`{=tex}\_2) and
(`\epsilon`{=tex}\_H`\ll1`{=tex}), \[
`\frac{z''}{z\tau^{-2}}`{=tex}`\simeq
2`{=tex}+`\frac32`{=tex}`\epsilon`{=tex}\_2+`\frac14`{=tex}`\epsilon`{=tex}\_2\^2.
\] This has the exact quadratic symmetry \[
`\boxed{\epsilon_2\longrightarrow -6-\epsilon_2}`{=tex}, \] which
exchanges slow roll, (`\epsilon`{=tex}\_2`\simeq0`{=tex}), with
ultra-slow roll, (`\epsilon`{=tex}\_2`\simeq-6`{=tex}).

The two independent (k`\rightarrow0`{=tex}) solutions, \[
{`\cal `{=tex}R}=C_1+C_2`\int`{=tex}`\frac{d\tau}{z^2}`{=tex}, \] then
behave completely differently: slow roll gives a decaying mode, whereas
USR gives a growing mode.

The possible new result is to show that this is the lowest member of a
more general **flow-space involution**, and derive its action on the
higher hierarchy.

My next move would be to do that derivation rather than produce prose.

------------------------------------------------------------------------
\end{Transcript}

\section*{User}
\begin{Transcript}
Sounds interesting, see what you can come up with. Feel free
to take your time and deploy agents as necessary.
\end{Transcript}

\section*{Assistant}
\begin{Transcript}
I'll treat this as a real research exercise rather than a
writing exercise. I'm going to try to either prove the flow-space
duality or kill it, then check whatever survives against the modern
constant-roll/USR literature before deciding whether there's a PRL-sized
result.

A useful correction has emerged. The naive
(`\eta`{=tex}\_H`\to3`{=tex}-`\eta`{=tex}\_H) symmetry is **not** an
exact symmetry of the full Hubble-flow hierarchy; the recent
constant-roll literature explicitly shows it fails once
(`\epsilon`{=tex}\_H) matters. But there is a deeper exact object behind
the approximation: the pump-field transformation that leaves (z''/z)
invariant and exchanges the two superhorizon solutions.

There is now a cleaner candidate than the original "full hierarchy
involution." In the (`\epsilon`{=tex}\_1`\to0`{=tex}) regime, demanding
**exactly constant Mukhanov--Sasaki mass through the entire
non-attractor->attractor transition** turns the flow equation into a
solvable Riccati equation. Its regular solution is a unique
(`\tanh`{=tex}) interpolation between the two dual constant-roll
branches. For the scale-invariant case it connects USR
((`\epsilon`{=tex}\_2=-6)) continuously to slow roll
((`\epsilon`{=tex}\_2=0)) while keeping (z''/z=2/`\tau`{=tex}\^2)
throughout.

The algebra has simplified in a satisfying way. For the scale-invariant
case, the unique regular Wands-preserving profile is exactly the general
solution for a field on a **linear potential** in the quasi-de Sitter
limit. The "USR branch" is the decaying homogeneous solution; the
"slow-roll branch" is the particular terminal-velocity solution.

I think there is something here, although I would **not yet claim PRL
novelty**. The original "full flow-hierarchy involution" does not
survive scrutiny. But following it led to a cleaner result tying
together old non-slow-roll/Hamilton--Jacobi work, constant roll, and
current USR/PBH work.

In the quasi-de Sitter limit, \[ `\mu`{=tex}\^2=
2+`\frac32`{=tex}`\epsilon`{=tex}\_2+`\frac14`{=tex}`\epsilon`{=tex}\_2\^2+`\frac12`{=tex}`\frac{d\epsilon_2}{dN}`{=tex}.
\] Demanding constant (`\mu`{=tex}\^2) gives \[
2`\epsilon`{=tex}\_2'+`\epsilon`{=tex}\_2\^2+6`\epsilon`{=tex}\_2+8-4`\mu`{=tex}\^2=0.
\] With (s=`\sqrt{1+4\mu^2}`{=tex}) and (y=`\epsilon`{=tex}\_2+3), \[
2y'=s^2-y^2, \] whose regular solution is \[
`\boxed{\epsilon_2(N)=-3+s\tanh\left[\frac{s}{2}(N-N_0)\right].}`{=tex}
\]

For scale invariance, (s=3): \[
`\boxed{\epsilon_2(N)=-3+3\tanh\left[\frac32(N-N_0)\right],}`{=tex} \]
which runs continuously from USR to SR while (z''/z=2/`\tau`{=tex}\^2)
remains unchanged.

Integrating gives \[
`\epsilon`{=tex}*1(N)=`\epsilon`{=tex}*{`\rm SR`{=tex}}`\left[1+e^{-3(N-N_0)}\right]`{=tex}\^2,
\] so \[
`\dot`{=tex}`\phi`{=tex}=`\dot`{=tex}`\phi`{=tex}*{`\rm SR`{=tex}}`\left[1+e^{-3(N-N_0)}\right]`{=tex}.
\] Then \[
`\ddot`{=tex}`\phi`{=tex}+3H`\dot`{=tex}`\phi`{=tex}=3H`\dot`{=tex}`\phi`{=tex}*{`\rm SR`{=tex}}=`\mathrm{constant}`{=tex},
\] and the Klein--Gordon equation requires
(V\_{,`\phi`{=tex}}=`\mathrm{constant}`{=tex}): a linear potential.

The scale-invariant case is the (m\^2=0) member of a larger statement.
For a locally quadratic potential with approximately constant (H), \[
`\ddot`{=tex}`\phi`{=tex}+3H`\dot`{=tex}`\phi`{=tex}+m\^2`\phi=0`{=tex},
\] the characteristic roots obey \[
`\lambda`{=tex}\^2+3`\lambda`{=tex}+`\frac{m^2}{H^2}`{=tex}=0, \] and
therefore the two constant-roll values satisfy \[
`\eta`{=tex}*++`\eta`{=tex}*-=3. \] So the constant-roll duality follows
from Vieta's theorem applied to the Klein--Gordon characteristic
equation.

The ingredients separately are known, however, and I would not yet write
the paper. The make-or-break direction is finite-(`\epsilon`{=tex}\_1)
breaking.

------------------------------------------------------------------------
\end{Transcript}

\section*{User}
\begin{Transcript}
Keep going.
\end{Transcript}

\section*{Assistant}
\begin{Transcript}
I'm pushing on the finite-(`\epsilon`{=tex}\_1) system
now. The main question is whether the quasi-de Sitter heteroclinic
solution deforms into an exact invariant curve, or whether finite
gravitational backreaction necessarily destroys constant (z''/z).

There is an important finite-(`\epsilon`{=tex}) subtlety: holding
(z''/\[z(aH)\^2\]) constant is **not** the same as holding the Hankel
index constant once (`\epsilon`{=tex}\_1`\neq0`{=tex}), because
(aH`\neq-1`{=tex}/`\tau`{=tex}). So the exact problem is
three-dimensional.

One candidate is now substantially **pre-empted**. Karam et al. (2023)
explicitly use the same (`\tanh`{=tex})-type Wands-preserving USR->CR
interpolation and realize it with quadratic potentials; and your 2018
paper already gives the "two roots of the background equation"
interpretation.

What does look cleaner is an exact finite-(`\epsilon`{=tex}) statement:
exact Wands-isospectral evolution can be written as a three-dimensional
autonomous flow, and the non-attractor/attractor heteroclinic orbit is
forced onto the (`\epsilon=0`{=tex}) boundary. In other words, finite
kinetic energy necessarily breaks the exact duality.

For \[ q`\equiv`{=tex}-aH`\tau`{=tex}, \] one has \[
q'=(1-`\epsilon`{=tex}\_1)q-1. \] Demanding \[
`\frac{z''}{z}`{=tex}=`\frac{C}{\tau^2}`{=tex} \] gives the exact system
\[ `\epsilon`{=tex}\_1'=`\epsilon`{=tex}\_1`\epsilon`{=tex}\_2, \] \[
`\epsilon`{=tex}\_2'= `\frac{2C}{q^2}`{=tex}-4+2`\epsilon`{=tex}\_1
-3`\epsilon`{=tex}\_2-`\frac12`{=tex}`\epsilon`{=tex}\_2\^2+`\epsilon`{=tex}\_1`\epsilon`{=tex}\_2,
\] \[ q'=(1-`\epsilon`{=tex}\_1)q-1. \]

On the quasi-de Sitter boundary (`\epsilon`{=tex}*1=0,q=1), the two
fixed points obey \[ `\epsilon`{=tex}*{2,+}+`\epsilon`{=tex}\_{2,-}=-6.
\]

If an exact isospectral trajectory approaches the non-attractor fixed
point in the asymptotic past, then \[
`\epsilon`{=tex}*1(N)`\sim `{=tex}`\epsilon`{=tex}*1(N_0)e\^{`\epsilon`{=tex}*{2,-}(N-N_0)}.
\] Because (`\epsilon`{=tex}*{2,-}\<0), the only trajectory with
(`\epsilon`{=tex}\_1`\to0`{=tex}) in the past is
(`\epsilon`{=tex}\_1`\equiv0`{=tex}). Thus an exact Wands-dual
heteroclinic orbit connecting the two constant-roll fixed points exists
only on the (`\epsilon`{=tex}\_1=0) boundary.

There is also a field-range result. During ideal USR, \[
`\epsilon`{=tex}*2`\simeq-6`{=tex}, \] and the field excursion is
bounded by \[
\|`\Delta`{=tex}`\phi`{=tex}*{`\rm USR`{=tex}}\|\<`\frac{\sqrt2}{3}`{=tex}M\_{`\rm P`{=tex}}.
\]

This suggests a possible framing: **An inverse Lyth bound for
ultra-slow-roll inflation.**

------------------------------------------------------------------------
\end{Transcript}

\section*{User}
\begin{Transcript}
Keep going.
\end{Transcript}

\section*{Assistant}
\begin{Transcript}
I'm generalizing the no-go statement away from constant
Hankel index now.

A potentially stronger connection surfaced from the loop/PBH literature,
but the exact finite-(`\epsilon`{=tex}) route is becoming less
attractive as a PRL by itself. The finite-(`\epsilon`{=tex}) correction
is real, but likely too incremental unless it produces a sharp lower
bound on non-Gaussianity or loops. I'm keeping that calculation but
opening a second line rather than forcing it.

I found a much cleaner exact statement that does **not** need the
(`\epsilon`{=tex}`\ll1`{=tex}) expansion.

For constant Hankel index, \[
z(`\tau`{=tex})=A(-`\tau`{=tex})^{1/2-`\nu`{=tex}}+B(-`\tau`{=tex})^{1/2+`\nu`{=tex}}.
\] For every observationally relevant (`\nu`{=tex}\>1/2), the second
branch grows without bound toward the asymptotic past. But canonical
inflation has \[
z=a`\sqrt{2\epsilon}`{=tex},M\_{`\rm P`{=tex}},`\qquad 0`{=tex}`\le`{=tex}`\epsilon`{=tex}\<1,
\] hence \[ \|z\|\<`\sqrt`{=tex}2M\_{`\rm P`{=tex}}a. \] An expanding
inflationary solution cannot support that past-growing branch
indefinitely. So the exact Wands-dual non-attractor phase is necessarily
finite-duration.

There is another corollary that is more general than the USR estimate. A
canonical phase has a genuinely anti-damped superhorizon mode when \[
3+`\epsilon`{=tex}*2\<0. \] If this holds throughout the interval, then
\[ `\epsilon`{=tex}\<`\epsilon`{=tex}*{`\rm in`{=tex}}e\^{-3N}, \] and
\[ `\frac{|\Delta\phi|}{M_{\rm P}}`{=tex}
=`\int`{=tex}`\sqrt{2\epsilon}`{=tex},dN \<
`\frac{2\sqrt2}{3}`{=tex}`\sqrt{\epsilon_{\rm in}}`{=tex} \<
`\frac{2\sqrt2}{3}`{=tex}. \] Thus every canonical single-field phase
with sustained superhorizon anti-damping is necessarily sub-Planckian.

For (`\epsilon`{=tex}\_2=-p), (p\>3), \[
`\frac{|\Delta\phi|}{M_{\rm P}}`{=tex} =
`\frac{2\sqrt{2\epsilon_i}}{p}`{=tex}
`\left`{=tex}(1-e\^{-p`\Delta `{=tex}N/2}`\right`{=tex}). \] The scalar
amplification behaves approximately as \[
{`\cal `{=tex}A}`\simeq `{=tex}e\^{2(p-3)`\Delta `{=tex}N}, \] so \[
`\frac{|\Delta\phi|}{M_{\rm P}}`{=tex} `\simeq`{=tex}
`\frac{2\sqrt{2\epsilon_i}}{p}`{=tex}
`\left[1-{\cal A}^{-\frac{p}{4(p-3)}}\right]`{=tex}. \] For USR, \[
`\frac{|\Delta\phi|}{M_{\rm P}}`{=tex} `\simeq`{=tex}
`\frac{\sqrt{2\epsilon_i}}`{=tex}3(1-{`\cal `{=tex}A}\^{-1/2}). \]

This is a **non-attractor Lyth relation**: unlike the ordinary Lyth
relation, increasing scalar amplification eventually costs essentially
no additional field distance.

I now have two live candidates: the inverse-Lyth theorem and a
BGV/stochastic-inflation project.

------------------------------------------------------------------------
\end{Transcript}

\section*{User}
\begin{Transcript}
Do the inverse Lyth bound. Keep going until you produce a
draft paper, even if you think it it insufficiently novel, as long as
it's correct.
\end{Transcript}

\section*{Assistant}
\begin{Transcript}
I'm going to rebuild the bound from the **exact**
superhorizon equation before writing anything. One correction is already
necessary: at finite (`\epsilon`{=tex}), care is needed in translating
between growth criteria in different variables. I'll formulate the
theorem operationally using the exact superhorizon equation.

The core result survived the exact rewrite. For a canonical scalar, \[
`\frac{d}{dt}`{=tex}(a\^3`\epsilon`{=tex},`\dot{\cal R}`{=tex})=0. \]
When \[ 3+`\epsilon`{=tex}\_2\<0, \] the nonconstant mode is
anti-damped. That condition alone gives the field-range upper bound; no
slow-roll or flat-potential approximation is needed.

I then produced a full draft titled:

**An Inverse Lyth Bound for Non-Attractor Inflation**

The principal theorem was \[ `\boxed{
\frac{\Delta\phi_{\rm NA}}{M_{\rm P}}
<
\frac{2\sqrt{2\epsilon_{\rm in}}}{3}
\left(1-e^{-3\Delta N/2}\right)
<
\frac{2\sqrt{2\epsilon_{\rm in}}}{3}.
}`{=tex} \] The weaker universal inflationary maximum is \[
`\frac{\Delta\phi_{\rm NA}}{M_{\rm P}}`{=tex}\<`\frac{2\sqrt2}{3}`{=tex}`\simeq0.94`{=tex}.
\]

For (`\epsilon`{=tex}\_2=-p\<-3), \[ `\boxed{
\frac{\Delta\phi}{M_{\rm P}}
=
\frac{2\sqrt{2\epsilon_{\rm in}}}{p}
\left(1-e^{-p\Delta N/2}\right)
<
\frac{2\sqrt{2\epsilon_{\rm in}}}{p}.
}`{=tex} \]

For USR, \[ `\boxed{
\frac{\Delta\phi_{\rm USR}}{M_{\rm P}}
=
\frac{\sqrt{2\epsilon_{\rm in}}}{3}
\left(1-e^{-3\Delta N}\right)
<
\frac{\sqrt{2\epsilon_{\rm in}}}{3}.
}`{=tex} \]

Defining exact velocity amplification \[
{`\cal `{=tex}G}`\equiv`{=tex}`\left`{=tex}\|`\frac{\dot{\cal R}_{\rm f}}{\dot{\cal R}_{\rm in}}`{=tex}`\right`{=tex}\|
=e\^{(p-3)`\Delta `{=tex}N}, \] gives the exact inverse-Lyth relation \[
`\boxed{
\frac{\Delta\phi}{M_{\rm P}}
=
\frac{2\sqrt{2\epsilon_{\rm in}}}{p}
\left[
1-{\cal G}^{-\frac{p}{2(p-3)}}
\right].
}`{=tex} \]

In the quasi-de Sitter limit, when the growing contribution dominates,
\[ {`\cal `{=tex}A}`\equiv`{=tex}
`\frac{{\cal P}_{{\cal R},f}}{{\cal P}_{{\cal R},i}}`{=tex}
`\simeq `{=tex}e\^{2(p-3)`\Delta `{=tex}N}, \] so \[ `\boxed{
\frac{\Delta\phi}{M_{\rm P}}
\simeq
\frac{2\sqrt{2\epsilon_{\rm in}}}{p}
\left[
1-{\cal A}^{-\frac{p}{4(p-3)}}
\right].
}`{=tex} \]

I concluded that the (`\epsilon`{=tex}*{`\rm in`{=tex}})-dependent bound
should be the headline theorem, rather than the (0.94M*{`\rm P`{=tex}})
number, and that the finite-(`\epsilon`{=tex}) power-spectrum relation
should be treated carefully.

------------------------------------------------------------------------
\end{Transcript}

\section*{User}
\begin{Transcript}
Make the edits you suggested and give me a downloadable file
in LaTeX format. It's okay if the length runs over a bit.
\end{Transcript}

\section*{Assistant}
\begin{Transcript}
I created a revised LaTeX draft. The two substantive
changes were:

1.  The \*\*(`\epsilon`{=tex}\_{`\rm in`{=tex}})-dependent
    inequality\*\* was made explicitly the headline theorem;
    (2`\sqrt2`{=tex}/3) appears only as the weaker absolute maximum.
2.  The finite-(`\epsilon`{=tex}) amplification was treated carefully.
    The relation to (`\dot{\mathcal R}`{=tex}) is exact, while the
    relation to scalar power is identified as quasi-de Sitter and
    dependent on the growing mode dominating. The exact
    finite-(`\epsilon`{=tex}) quadrature for (`\mathcal `{=tex}R) was
    included.

The file was provided as `inverse_lyth_bound.tex`.

------------------------------------------------------------------------
\end{Transcript}

\section*{User}
\begin{Transcript}
Could you please reformat the bibliography in BibTeX format,
and give me the paper as two separate files, paper.tex and
bibliography.bib.
\end{Transcript}

\section*{Assistant}
\begin{Transcript}
Done. The paper now uses `apsrev4-2` with the
bibliography in a separate BibTeX database. I provided `paper.tex` and
`bibliography.bib`.

------------------------------------------------------------------------
\end{Transcript}

\section*{User}
\begin{Transcript}
I want to make it clear that the paper is AI-generated, and
not my own work. Because I am an ethical person. Can you come up with an
author name that is an anagram of "Will Kinney" or "William Kinney" or
some suitable variation, and "ChatGPT" or "ChatGPT Sol" or some suitable
variation?
\end{Transcript}

\section*{Assistant}
\begin{Transcript}
Yes. That's a fun way to make the provenance obvious
while still giving the experiment a pseudonymous "author."

I suggested constructing a single plausible human-sounding pseudonym as
an exact anagram of

**WILLIAM KINNEY CHATGPT SOL**

with every letter used exactly once, and pairing it with an explicit
disclosure such as:

> The author name is a pseudonym and an anagram of "William Kinney +
> ChatGPT Sol." This manuscript was generated by ChatGPT as an
> experiment in AI-assisted scientific research. William H. Kinney did
> not author the manuscript.

------------------------------------------------------------------------
\end{Transcript}

\section*{User}
\begin{Transcript}
Can you make it more anagram-ey?
\end{Transcript}

\section*{Assistant}
\begin{Transcript}
Yes. I proposed making the constraint exact:

\[ `\boxed{\texttt{WILLIAM KINNEY CHATGPT SOL}}`{=tex} \]

with spaces, punctuation, capitalization, and perhaps middle initials
free, but every letter used exactly once.

------------------------------------------------------------------------
\end{Transcript}

\section*{User}
\begin{Transcript}
Use that.
\end{Transcript}

\section*{Assistant}
\begin{Transcript}
I agreed to solve it as a true anagram, with the letter
bank

**A A C E G H I I I K L L L M N N O P S T T W Y**

and optimize for a plausible human name.

------------------------------------------------------------------------
\end{Transcript}

\section*{User}
\begin{Transcript}
See what you can come up with.
\end{Transcript}

\section*{Assistant}
\begin{Transcript}
I treated it as a constrained naming problem: exact
letter conservation first, plausibility second.

I proposed several exact-anagram names, with my favorite:

**G. P. T. Wilkins Lacey Hamilton**

and suggested the disclosure:

> **G. P. T. Wilkins Lacey Hamilton**\
> *The author name is an exact anagram of "William Kinney ChatGPT Sol."
> This manuscript was generated by ChatGPT (GPT-5.6 Sol) as an
> experiment in AI-generated scientific research and was not authored by
> William H. Kinney.*

------------------------------------------------------------------------
\end{Transcript}

\section*{User}
\begin{Transcript}
Can you give me an anagram for "University at Buffalo, SUNY",
or similar?
\end{Transcript}

\section*{Assistant}
\begin{Transcript}
I noted that the letter inventory is awkward and
suggested using a fictional institutional name. One appealing direction
was **Albufny University**, retaining visible genetic residue from
*Buffalo + SUNY* while clearly being fictional.

------------------------------------------------------------------------
\end{Transcript}

\section*{User}
\begin{Transcript}
Go.
\end{Transcript}

\section*{Assistant}
\begin{Transcript}
I optimized for a "real-but-fictional institution." The
exact letter bank for

**UNIVERSITY AT BUFFALO SUNY**

is

**A A B E F F I I L N N O R S S T T U U U V Y Y**.

I suggested that **Albufny University** is aesthetically effective even
though an exact natural-language institutional anagram is difficult, and
that relaxing the source string to "University at Buffalo" or "SUNY
Buffalo" would permit cleaner exact anagrams.

------------------------------------------------------------------------
\end{Transcript}

\section*{User}
\begin{Transcript}
This is awesome, thanks! Can you give me a downloadable
transcript of this conversation?
\end{Transcript}

\end{document}